\documentclass[journal=nalefd,manuscript=article]{achemso}

\usepackage{chemformula} 
\usepackage[T1]{fontenc} 
\usepackage{float} 
\usepackage{xcolor} 
\usepackage[T1]{fontenc}
\usepackage[latin9]{inputenc}
\usepackage{color}
\usepackage{units}
\usepackage{amssymb}
\usepackage{graphicx}
\usepackage{esint}
\usepackage{bm}
\usepackage{natbib}
\usepackage{amsmath}

\usepackage{ulem}
\setcitestyle{journalcolor= blue}

\author{Priya Tiwari}
\affiliation{Department of Physics, Indian Institute of Science, Bangalore 560012, India}
\author{Saurabh Kumar Srivastav}
\affiliation{Department of Physics, Indian Institute of Science, Bangalore 560012, India}
\author{Sujay Ray}
\affiliation{Department of Physics, Indian Institute of Science, Bangalore 560012, India}
\author{Tanmoy Das}
\affiliation{Department of Physics, Indian Institute of Science, Bangalore 560012, India}

\author{Aveek Bid}
\email{aveek@iisc.ac.in}
\affiliation{Department of Physics, Indian Institute of Science, Bangalore 560012, India}

\title[An \textsf{achemso} demo]{Observation of Time-Reversal Invariant Helical Edge-Modes in Bilayer Graphene/\ch{WSe2} Heterostructure}


\begin{document} 
	
\begin{abstract}
		
Topological insulators, along with Chern insulators and Quantum Hall insulator phases, are considered as paradigms for symmetry protected topological phases of matter. This article reports the experimental realization of the time-reversal invariant helical edge-modes in bilayer graphene/monolayer \ch{WSe2}-based heterostructures -- a phase generally considered as a precursor to the field of generic topological insulators. Our observation of this elusive phase depended crucially on our ability to create mesoscopic devices comprising both a moir\'e superlattice potential and strong spin-orbit coupling; this resulted in materials whose electronic band structure could be tuned from trivial to topological by an external displacement field. We find that the topological phase is characterized by a bulk bandgap and by helical edge-modes with electrical conductance quantized exactly to $2e^2/h$ in zero external magnetic field. We put the helical edge-modes on firm grounds through supporting experiments, including the verification of predictions of the Landauer-B$\mathrm{\ddot{u}}$ttiker model for quantum transport in multi-terminal mesoscopic devices. Our non-local transport properties measurements show that the helical edge-modes are dissipationless and equilibrate at the contact probes. We achieved the tunability of the different topological phases with electric and magnetic fields, which allowed us to achieve topological phase transitions between trivial and multiple, distinct topological phases. We also present results of a theoretical study of a realistic model which, in addition to replicating our experimental results, explains the origin of the topological insulating bulk and helical edge-modes. Our experimental and theoretical results establish a viable route to realizing the  time-reversal invariant $\mathbb{Z}_2$ topological phase of matter.

\end{abstract}

\section{Introduction}
Topological phases of matter are characterized by specific global properties that emerge from symmetry protected, local degrees of freedom~\cite{RevModPhys.82.3045, RevModPhys.88.021004, Ren_2016}; prominent examples being Chern insulators and $\mathbb{Z}_2$ invariant topological insulators. Chern insulators are phases with broken time-reversal (TR) symmetry which host chiral edge-modes~\cite{weng2015quantum} -- examples being the Quantum Hall and the Anomalous quantum Hall phases. The TR invariant $\mathbb{Z}_2$ topological phase, on the other hand, has counter-propagating helical edge-modes. There are two distinct proposals for the realization of a TR invariant topological phase in two-spatial dimensions. The proposal by Kane and Mele~\cite{PhysRevLett.95.146802, PhysRevLett.95.226801} predicted Quantum Spin Hall insulator (QSHI) phase in a carbon honeycomb lattice with strong spin-orbit coupling (SOC). In this model, the separation between the helical edge-modes arises from the SOC -- the spin-$\uparrow$ and spin-$\downarrow$ charge-carriers experience opposite forces due to SOC and form edge-modes of opposite chirality. Alternately, Bernevig and Zhang considered strained Zinc-Blende semiconductors where the U(1) symmetry is broken either due to bulk asymmetry or Rashba spin-orbit coupling leading to the QSHI phase~\cite{PhysRevLett.96.106802,bernevig2006quantum}. Although the QSHI phase has experimentally been realized in various systems, including HgTe/HgCdTe, InAs/GaSb~\cite{Knig766, PhysRevLett.107.136603} quantum wells and WTe$_2$~\cite{Wu76}, its confirmation in carbon honeycomb lattice is lacking. 

The principal impediments to the presence of QSHI phase in graphene are (i) lack of a bandgap and, (ii) the small value of its intrinsic SOC ($\sim 2$~$ \mathrm{\mu eV}$).  In bilayer graphene (BLG), the presence of an external electric field perpendicular to the plane  lifts the $z\rightarrow -z$ inversion symmetry and creates a bandgap at the $K$ and $K'$ points. When such a gapped BLG is stacked on a material with strong SOC (\textit {e.g.}, monolayer transition metal dichalcogenides like \ch{WSe2}), a strong, layer-selective SOC of spin valley locking nature is inherited by the BLG from the \ch{WSe2}. Calculations have shown that this can, under appropriate conditions (which are discussed in detail later in this article) place the BLG in a topologically non-trivial phase~\cite{PhysRevLett.119.146401, Wang2015} which  hosts metallic helical edge-modes and forms a viable platform for the observation of  QSHI~\cite{PhysRevLett.107.256801, PhysRevB.94.241106, PhysRevB.92.155403, Yang_2016, PhysRevB.82.195438}. 
	
This article reports the experimental observation of TR invariant helical edge-modes in high-mobility heterostructures of BLG/monolayer \ch{WSe2}. The measured conductance in several multi-probe measurement configurations (both local and non-local) was found to be precisely quantized over a range of temperatures (20~mK - 10~K) {(see Fig. S4)}; the quantized value in all cases being equal to that expected for the helical edge structure of QSHI. The presence of the helical edge-modes is placed on firm grounds through the verification of predictions of the Landauer-B$\mathrm{\ddot{u}}$ttiker model~\cite{PhysRevB.38.9375}  for quantum transport in multi-terminal devices with helical edge-modes and through measurements of non-local transport properties. Through theoretical studies of a realistic model of our system, we accurately replicate our experimental results. From our coupled experimental and theoretical study, we establish that the necessary conditions for the observation of a pair of helical edge-modes in BLG are: (i) presence of a significant bulk bandgap, (ii) presence of asymmetric Ising spin-orbit interaction acting predominantly on one layer of the Bernal stacked BLG, and (iii) the presence of a moderately strong Rashba spin-orbit interaction that gaps out all low-energy edge-modes in the  BLG except at the $K$ and the $K'$ points.
	
\section{Results and Discussion}
\begin{figure*}[t]
		\includegraphics[width=\textwidth]{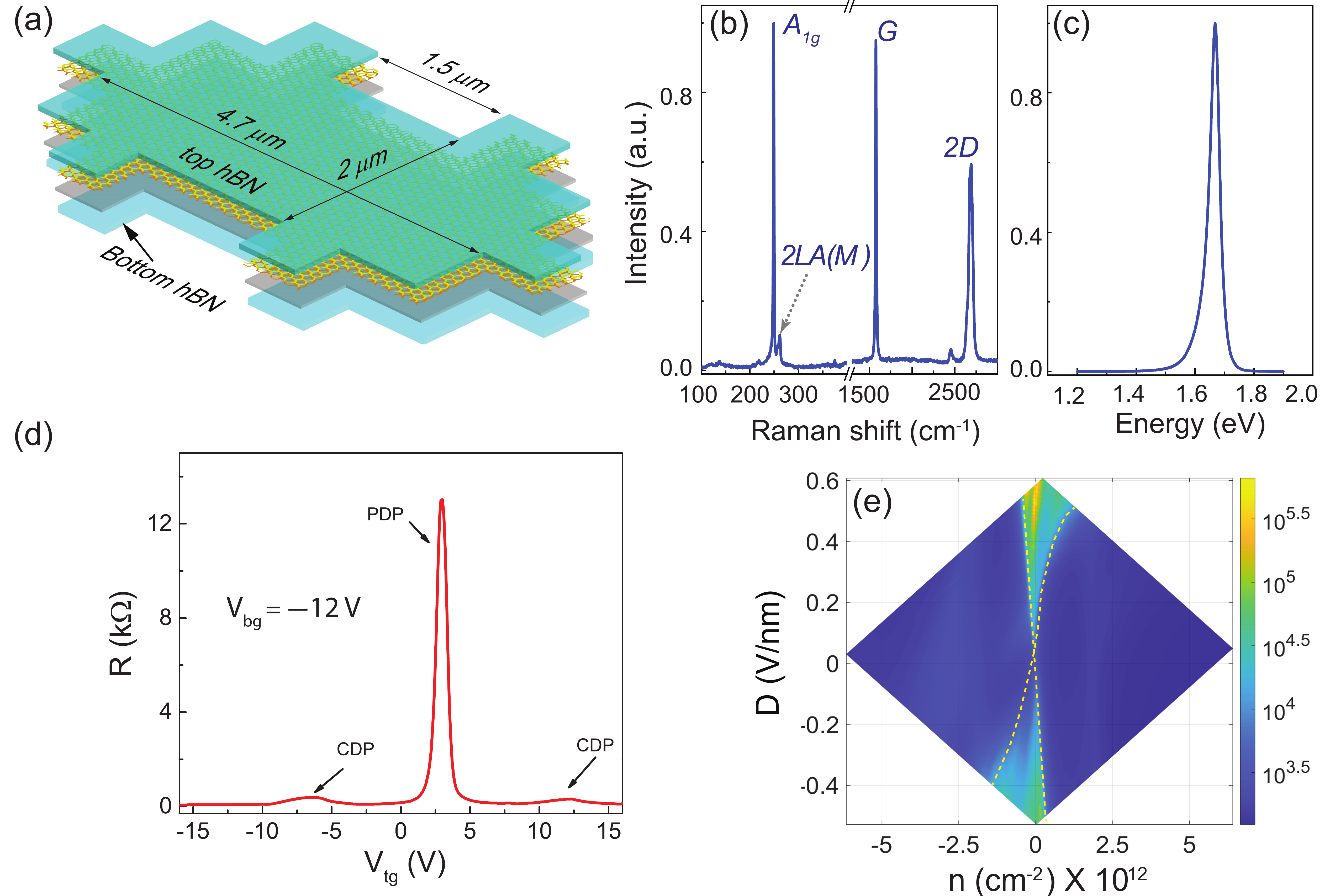}
		\small{\caption{ Device structure and characterization. (a) A schematic of the device configuration. The co-laminated heterostructure of BLG (shown as honeycombed structure) and monolayer \ch{WSe2} (grey layer) is sandwiched between two hBN flakes, each of thickness $\sim 20$~nm. (b) Room temperature Raman spectra of the \ch{WSe2} and BLG flakes. The peaks corresponding to monolayer \ch{WSe2} and BLG are marked.  (c) Room temperature photoluminescence response of \ch{WSe2} flake -- the peak at 1.65~eV establishes it to be a monolayer. (d) Four-probe resistance plotted as a function of $V_{tg}$ keeping back gate voltage fixed at $-12$~V; arrows mark the clone Dirac point (CDP) and primary Dirac point (PDP). (e) Logarithmic scale color plot of the 2-probe longitudinal resistance \textit {versus} the displacement field $\left|D\right|$ and the net charge carrier concentration, $n$. The dashed lines mark the asymmetry around the primary Dirac point. The data were acquired at $T=8$~K and $B=0$~T.  
				\label{fig:figure1}}}
	\end{figure*}

Hall bar devices ({B10S5 and B10S7}) based on an atomically sharp interface between a Bernal-stacked BLG and a monolayer of semiconducting crystalline \ch{WSe2} were fabricated using standard dry transfer technique~\cite{pizzocchero2016hot, Wang614}. The BLG/\ch{WSe2} stack was encapsulated between hexagonal boron nitride (hBN) crystals of thicknesses  $\sim$51~nm and $\sim$105~nm thicknesses at bottom and top of the stack (Fig.~\ref{fig:figure1}(a)) which acted as gate dielectrics for the back- and top-gates , respectively (for details of device fabrication see Supplementary Information (SI)). Fig. 1(b) is the Raman spectra of the BLG/\ch{WSe2} stack.  The two Raman peaks centered around 1580 cm$^{-1}$ (G peak) and 2800 cm$^{-1}$ (2D peak) belong to the graphene layer. The spectral decomposition of the 2D peak (see Fig. S1(C)), as well as the intensity ratio of G and 2D peaks, establish the graphene to be a bilayer~\cite{MALARD200951}. The presence of a peak near 250 cm$^{-1}$ (and the absence of one at around 308 cm$^{-1}$)\cite{C3NR03052K}  indicate that the \ch{WSe2} flake is a monolayer. Room temperature photoluminescence spectra of the \ch{WSe2} flake (Fig. 1(c)) has a peak at $\sim$~1.65 eV, confirming that it is a monolayer~\cite{Tonndorf:13}. All the data presented here are from device B10S5 -- the data from device B10S7 is presented  in supplementary material (see Fig. S5; SI6).

The dual-gated architecture of the device allows independent tuning of both the charge-carrier number density, $n$, and the displacement field perpendicular to the device, $D$, using the relation $n=((C_{tg}V_{tg}+C_{bg}V_{bg})/e)-n_o$ and $D=((C_{tg}V_{tg}-C_{bg}V_{bg})/2\epsilon_{o})-D_o$, where, $n_o$ is the residual charge-density due to doping and $D_o$  the net internal displacement field. The values of $n_o$ and $D_o$ were extracted from the position of the Dirac point in the $V_{tg}-V_{bg}$ plane. A plot of the four-probe longitudinal resistance as functions of the top-gate voltage, $V_{tg}$ at the back-gate voltage, $V_{bg}$ equal to $-12$~V is shown in Fig.~\ref{fig:figure1}(d). Arrows show the primary Dirac point (PDP) and the two satellite peaks, called clone Dirac points (CDP). These two satellite peaks are the result of band structure reconstruction of BLG due to the moir$\mathrm{\acute{e}}$ superlattice potential caused by a near-perfect alignment of the top hBN layer with the BLG~\cite{ponomarenko2013cloning,dean2013hofstadter, PhysRevB.14.2239}. From the positions of the CDP, the angle between top hBN and the BLG was estimated to be $\sim 0.9^\circ$ (see SI). In Fig.~\ref{fig:figure1}(e) is plotted a contour map of the 2-probe device resistance \textit {versus} $n$ and $D$. The asymmetric feature seen near the PDP (outlined by dashed lines) is a consequence of the band-splitting in the BLG by the induced SOC~\cite{island2019spin}. Depending on the direction of $D$,  this splitting occurs either in the conduction band (for positive values of $D$) or in the valence band (for negative values of $D$). Note that the appearance of the  moir$\mathrm{\acute{e}}$ potential induced subgaps in the band structure of the BLG ensures that the $K$ and the $K'$ points are no longer connected in the same band which  significantly suppresses inter-valley scattering in our system~\cite{Song10879}.

\subsection{Spin-orbit interaction in BLG}
	
\begin{figure*}	[t]
		\begin{center}
			\includegraphics[width=0.9\textwidth]{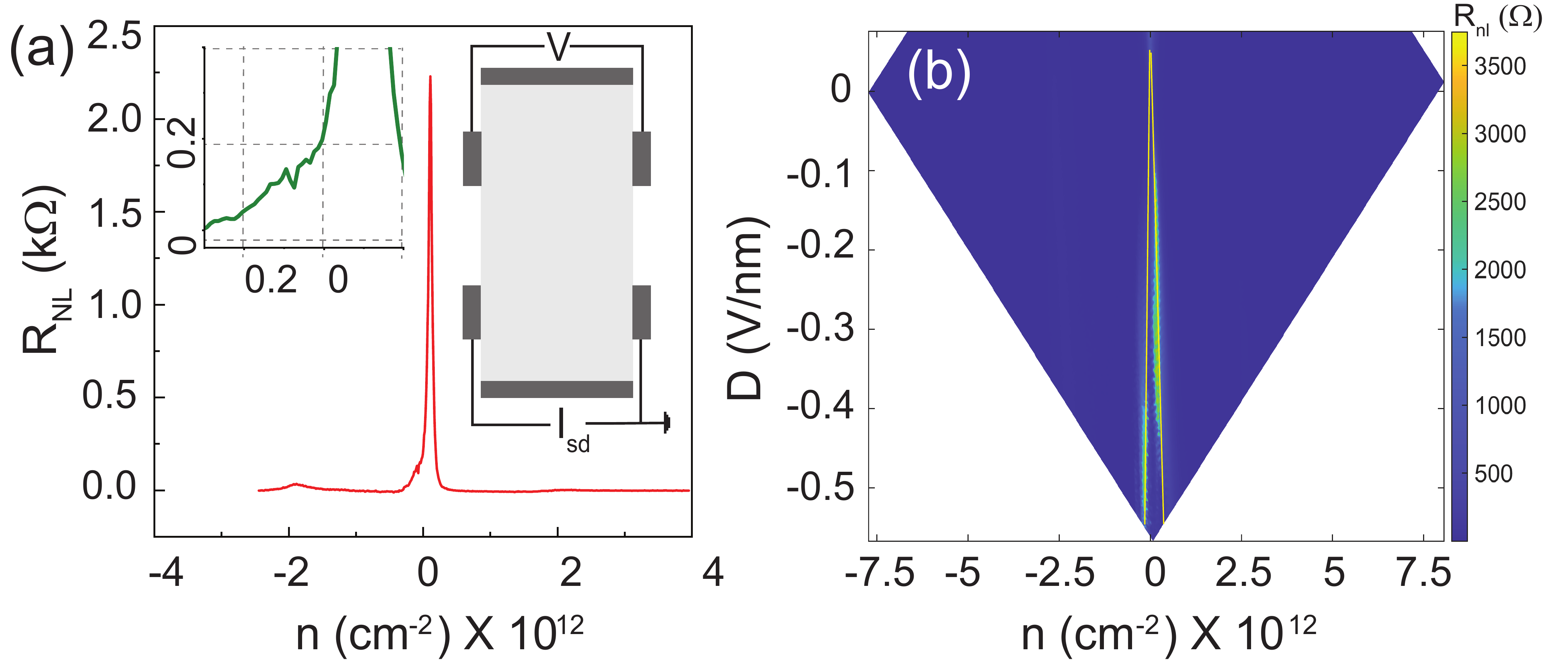}
			\small{\caption{(a) Plot of the non-local resistance, $R_{NL}$ \textit {versus} $n$ for $D =-0.098$~V/nm. Left inset: zoom-in of $R_{NL}$ near the PDP. Right inset: schematic of the measurement configuration. (b) Contour plot of $R_{NL}$ as a function of $n$ and $D$. The yellow dotted lines highlight the splitting observed in the $R_{NL}$ peak at large $\left|D\right|$. 
					\label{fig:figure2}}}
		\end{center}
	\end{figure*}

In a system with strong SOC, long-range spin currents can be generated in bulk by the spin Hall effect (SHE)~\cite{PhysRevLett.92.126603, Kato1910},  which produce a non-local signal at voltage probes remote from the charge current path \textit {via} the inverse SHE (for a schematic see Fig. S9, SI). The measured non-local resistance $R_{NL}$ in our device is peaked sharply near the primary Dirac point (Fig.~\ref{fig:figure2}(a)) and is at least three orders of magnitude larger than that expected from ohmic contributions alone (see SI). At large $\left|D\right|$, $R_{NL}$ has a split peak (Fig.~\ref{fig:figure2}(b)). Recall that the strength of the Berry curvature is most significant at the bulk band-edges, at the so-called `Berry curvature hot-spots.' This causes the effect of the SHE to be most prominent when the chemical potential of the system lies at one of the band-edges~\cite{Gorbachev448}. With increasing $\left|D\right|$, the bandgap in BLG increases causing the Berry curvature hot-spots in both the valence-band and the conduction-band to move apart -- this can be mapped out in our system by monitoring the location of the peak of $R_{NL}$ in the $n-D$ plane.

\subsection{Quantized transport through edge-modes}
	
\begin{figure*}	[t]
		\begin{center}
			\includegraphics[width= \textwidth]{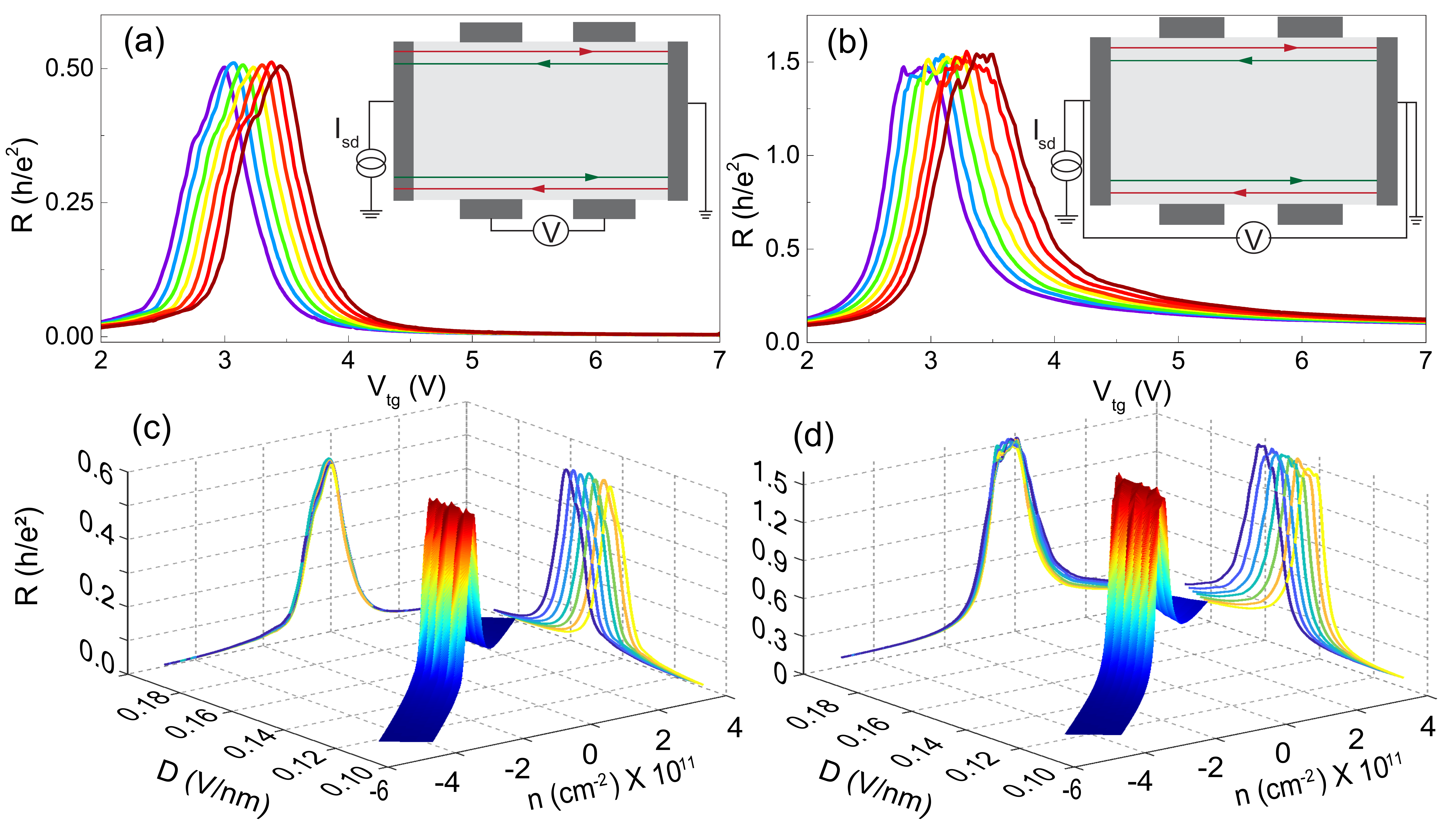}
			\small{\caption{(a) Quantization of the 4-probe resistance to $h/\left(2e^2\right)$ as $V_{tg}$ is varied at fixed values of $V_{bg}$ such that the Fermi level lies in the bulk bandgap. The value of $V_{bg}$ for the different plots varies from -12~V (brown line) to -13.5~V (dark blue line). The inset is a schematic of the 4-probe measurement configuration - the red and the green lines represent the spin-filtered edge-modes.  (b) Quantization of the 2-probe resistance to $3h/\left(2e^2\right)$, the data were acquired simultaneously with the data presented in (a). The inset is a schematic of the 2-probe measurement configuration (c) Plot of the 4-probe $R$ \textit {versus} $n$ and $\left|D\right|$. The projection on the $R-n$ plane shows that the quantization is around $n=0$, while the projection on the $R-D$ plane shows the quantization appears over a range of values of the displacement field. (d) Corresponding plot for 2-probe measurements -- the projection on the $R-n$ plane  shows  that the quantization is again at the PDP. All the data were acquired at $T = 8$~K (quantization obtained at 20~mK and for other cooldowns are presented in SI).
					\label{fig:figure3}}}
		\end{center}
	\end{figure*}

There are several experimental observations that confirm the appearance of a strong proximity induced SOC in our BLG system -- (i) presence of a substantial $R_{NL}$ signal at the band-edges, (ii) weak antilocalization correction to the low-field magnetoresistance (Fig. S6; SI3), and (iii) the asymmetry in the plot of $R\left(B=0\right)$ (Fig.~\ref{fig:figure1}(e)). This leads to the expectation that the system will host helical edge-modes.  (see SI 7). The plots of the measured 4-probe and 2-probe longitudinal resistances for B = 0 T and T = 8~K shown in Fig.~\ref{fig:figure3} establish that this indeed is the case. Fig.~\ref{fig:figure3}(a) shows precise quantization of the 4-probe longitudinal resistance to $h/\left(2e^2\right)$ at the PDP over a range of displacement fields. In Fig.~\ref{fig:figure3}(b) is plotted the measured 2-probe resistance which quantizes to $3h/\left(2e^2\right)$. These are the exact values one would expect for the respective measurement configurations for a system hosting an odd number of helical edge-modes (see SI for details). The quantization of the 2-probe and 4-probe longitudinal resistances at the PDP over a range of values of the displacement field is the central result of this article and establishes the emergence of helical edge-modes in our system. Fig.~\ref{fig:figure3}(c) and Fig.~\ref{fig:figure3}(d) are respectively the plots of the 4-probe, and 2-probe longitudinal resistances plotted \textit {versus} $n$ and $\left|D\right|$. The data projected on the $n-R$ plane show that the quantization in each case is centered around the PDP (with $\Delta n = \pm 2\times 10^{-15}$~m$^{-2}$).
	
\begin{figure*}	[t]
		\begin{center}
			\includegraphics[width= \textwidth]{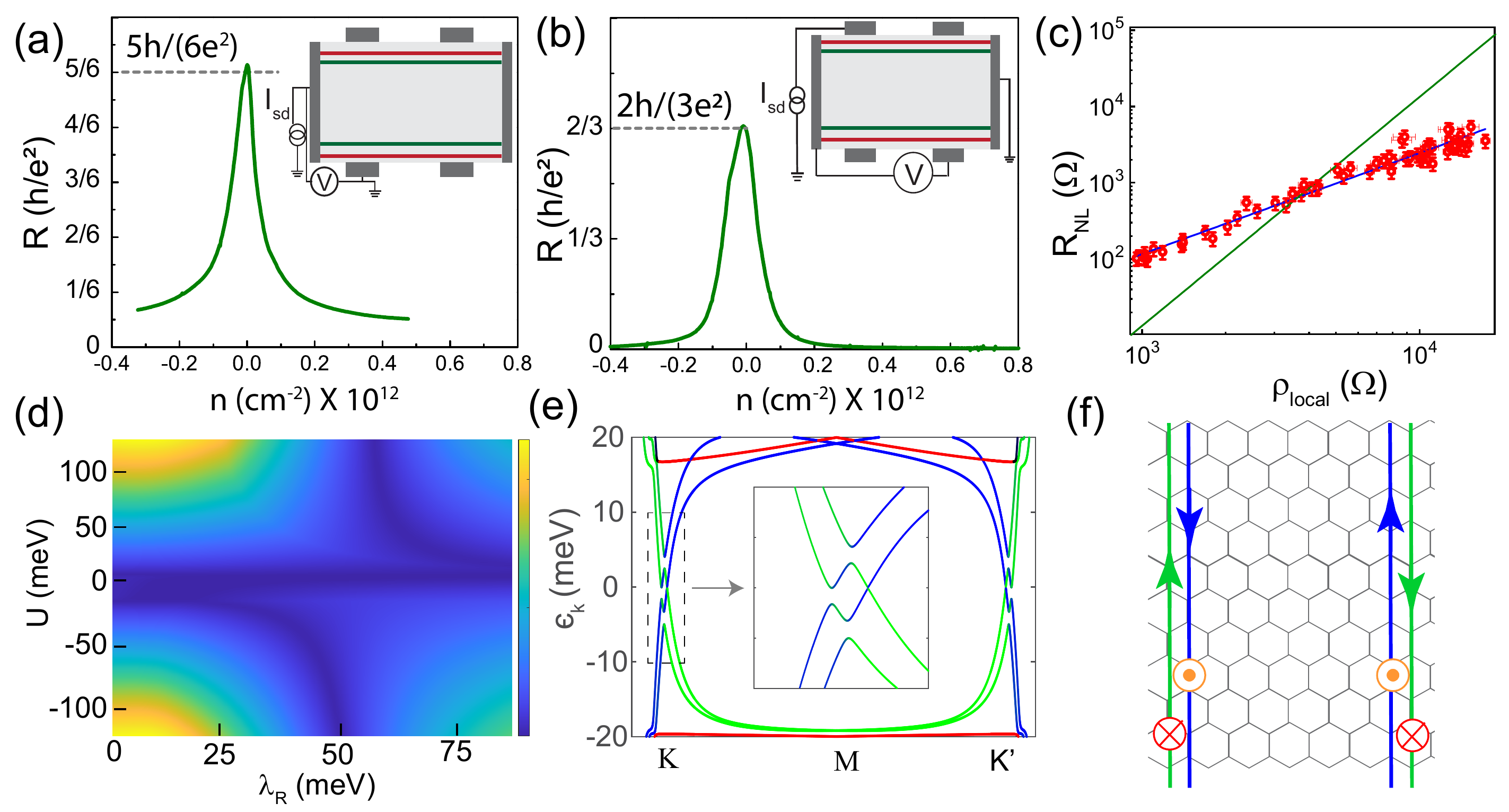}
			\small{\caption{(a) Quantization of $R$ to $2h/\left(3e^2\right)$ in a 3-probe measurement geometry - the measurement configuration is sketched in the inset. (b) Quantization of $R$ to $5h/\left(6e^2\right)$ - the inset shows the measurement configuration.  (c) Plot of $R_{NL}$ \textit {versus} the local longitudinal resistivity $\rho_{local}$ measured at different $D$. The filled red circles are the measured data points; the solid blue line is a fit to $R_{NL} \propto \rho_L$. The solid green line represents a plot of $R_{NL} \propto {\rho_L}^3$. (d) A bandgap phase diagram as a function of $U$ and $\lambda_{\rm R}$ plotted for representative values of  $\lambda_{ISO}^{B}=2$~meV and $\lambda_{ISO}^{T}=0.02$~meV.  The dark blue region marks the boundary between the topologically trivial and non-trivial regions of the phase diagram. (e) Emergence of helical edge-modes within the bulk bandgap  for $\lambda_{\rm R}=40$~meV and $U=-40$~meV. The inset shows a magnified view around the $K$ point of the edge-modes with a Dirac cone. (f) Schematic diagram showing the helical edge-modes arising from the band structure in panel (e). The blue and the green arrows represent respectively the counter-clockwise propagating and clockwise propagating helical edge modes; the red and the orange arrows indicate the direction of the out-of plane spin-component of these two modes.  
					\label{fig:figure4}}}
		\end{center}
	\end{figure*}

It might be argued that the observed quantization of the 4-probe longitudinal resistance can be due to the ballistic nature of the device or a fortuitous arrangement of scattering centers. To address these issues, the longitudinal resistance of the device was measured in several multi-terminal configurations. The data for two such configurations are shown in Fig.~\ref{fig:figure4}, along with the respective measurement configurations (data for additional configurations are presented in SI4). The gray lines are the values of longitudinal resistance calculated for the helical edge-modes using the Landauer-B$\mathrm{\ddot{u}}$ttiker formalism for quantum transport in multi-terminal devices. (Note that, for chiral edge-modes, the expected values of the resistance for the configuration in Fig.~\ref{fig:figure4}(a) would be zero and for that in Fig.~\ref{fig:figure4}(b) would be $h/e^2$). The excellent match between the measured resistance and the predictions based on helical edge-modes establishes that these edge-modes are dissipationless and that they equilibrate at the contact probes. This provides further confirmation that the system is most likely in the QSHI phase near the PDP. In the SI, we show that the two-probe resistance measured over two different length scales (1.5~$\mu$m and 5~$\mu$m) quantize to the expected values. The fact that almost identical quantization results were obtained over several thermal cycles of the device from room temperature to 20~mK (comparison of results from two different cool-downs of the device are presented in Fig. S3 of SI) also establishes the robustness of our results.
	
The $R_{NL}$ signal was measured as a function of the magnetic field, $B$ applied perpendicular to the plane of the device. Breaking of TR symmetry by the magnetic field would lead to a gradual quenching of the QSHI phase and consequent decay of the non-local signal -- the non-local signal can only persist as long as $\Delta_z>E_Z$, where $\Delta_z$ is the energy-gap for the QSHI phase and $E_Z$ is the Zeeman energy~\cite{Komatsueaaq0194, PhysRevB.81.195431}.  From the temperature dependence of the quantized conductance, $\Delta_z$ was estimated to be about 7~meV. This yields the maximum magnetic field range, till which the quantization can persist to be $B_{max}\sim \Delta_z^2/(m\mu_B v_F^2) = 0.1$~T, which matches very well with our observations (see SI).  Here $m$ is the mass of the electron, $\mu_B$ is the Bohr magneton and $v_F \sim 1.2\times{10}^6$ ms$^{-1}$ is the nominal Fermi velocity of the charge carriers in BLG. 
	
To obtain a microscopic understanding of the topological features of our setup, we use a tight-binding model with both Ising and Rashba SOCs for the AB-stacked BLG on \ch{WSe2} with an applied electric field $\left|D\right|\ \neq0$. The electric field plays several critical roles in this system: (a) It lifts the layer degeneracy in the BLG and opens a bandgap at the Fermi level.  (b) Above a certain critical value of the electric field, the top and bottom layers of the BLG  become fully charge polarized. In such a scenario, the conduction and the valence bands are formed of orbitals localized respectively in the negatively and positively charged layers of the BLG \cite{PhysRevLett.119.146401}. This leads to a strong correlation between the bands and the layers of the BLG.  (c) This layer-selective band structure receives SOC differently from the \ch{WSe2} layer. As the \ch{WSe2} is proximate to the bottom layer, it induces a much stronger SOC to the states localized at this layer layer, compared to that in the top layer. This leads to large ($\sim 2$~meV) spin-splitting of the conduction (valence) band for positive (negative) values of $D$.

To account for the above properties, we use the following model Hamiltonian for our system:
	\begin{eqnarray}
		H_{BLG}&=&H_{SLG}^{\rm T}+H_{SLG}^{\rm B}+t_{\perp}\sum_{i\in T,j\in B,\alpha}\left( c^{\dagger}_{i\alpha}c_{j\alpha} + c^{\dagger}_{j\alpha}c_{i\alpha}\right) \nonumber \\
		&&+\frac{U}{2}\sum_{i \in T,\alpha}c^{\dagger}_{i\alpha}c_{i\alpha}-\frac{U}{2}\sum_{i \in B,\alpha}c^{\dagger}_{i\alpha}c_{i\alpha}
	\end{eqnarray}
where, $H_{SLG}^{T/B}$ is the single-layer Hamiltonian -- the superscripts $\mathrm{T}$ and $\mathrm{B}$ refer to the top and bottom layers, respectively. $t_{\perp}$ is the inter-layer nearest-neighbor hopping amplitude. $U$ is the inter-layer onsite potential difference which includes contribution from: (a) asymmetry between the two layers due to proximity to the \ch{WSe2} layer (minor contribution) and, (b)  the applied electric field (larger contribution). For the single layer Hamiltonian, we include layer-specific SOC contributions from both  Ising SOC and Rashba SOC~\cite{PhysRevLett.107.256801}:
	\begin{eqnarray}
		H_{SLG}^{T/B}&=&-t\sum_{<ij>\alpha}c^{\dagger}_{i\alpha}c_{j\alpha}+i\lambda_{ISO}^{T/B}\sum_{<<ij>>\alpha}\nu_{ij}c^{\dagger}_{i\alpha}\left( s_{\alpha\alpha}\right)_{z}c_{j\alpha} \nonumber \\
		&& + it_{R}^{T/B}\sum_{<ij>\alpha\beta}\left( s_{\alpha\beta}\times {\bf d}_{ij}\right)_{z}c^{\dagger}_{i\alpha}c_{j\beta}
	\end{eqnarray}
where, $t$ is nearest neighbor in-plane hopping, $\lambda_{ISO}^{T/B}$ is the Ising SOC parameter, and $\lambda_{R}^{T/B}$ is the Rashba SOC parameter. The bandgap near the Fermi levels is primarily controlled by the competition between $U$ and  $\lambda_{\rm R}$, (the Ising SOC is somewhat less effective here). Therefore, we present a band-gap phase diagram as a function of $U$ and $\lambda_{\rm R}$ in Fig.~\ref{fig:figure4}(d) for a representative values of  $\lambda_{ISO}^{B}=2$~meV , $\lambda_{ISO}^{T}=0.02$~meV\cite{PhysRevLett.119.146401}. We find, consistent with previous calculations~\cite{PhysRevLett.107.256801}, that for small $\lambda_{\rm R}$, the band gap opens in the trivial side (yellow regions) and there is a band inversion (boundary marked by dark blue regions) to the non-trivial region as $\lambda_{\rm R}$ increases.

To evaluate the nature of the edge states in the non-trivial topological phase, we calculate the edge state profile with an open boundary condition along the zigzag direction and periodic along the other direction (see SI).  In Fig.~\ref{fig:figure4}(e) we present the edge state profile extracted from our calculations at representative values of  $\lambda_{\rm R}=40$~meV, and $U=-40$ meV. Indeed we find helical edge-modes emerging within the bulk bandgap with a Dirac cone -- a representative sketch is shown in Fig.~\ref{fig:figure4}(f) for the interpreted helical edge states. This result is consistent with our experimental observations. Note that in the absence of  Rashba SOC, we do not find any edge-modes. This is in contrast to the case of symmetric device structure (\ch{WSe2}/BLG/\ch{WSe2}) reported before where there are two pairs of helical edge-modes arising from Quantum Valley Hall effect giving rise to an edge conductance of $4e^2/h$ at $\lambda_{\rm R} \rightarrow 0$ ~\cite{PhysRevLett.107.256801} .
	
To estimate the strength of the Rashba SOC in our system, we assume that the Elliot Yafet mechanism is the dominant source of spin scattering in graphene~\cite{ochoa2012elliot} . In that case, the strength of the Rashba SOC is given by $\lambda_r = {2E_F}\sqrt{\frac{\tau_P}{\tau_{asy}}}$, where $E_F$ is the average Fermi energy, $\tau_{asy}$ is time scale corresponding to the spin-flip processes that breaks the $z\longrightarrow-z$ symmetry and $\tau_{p}^{-1}$ is the momentum relaxation rate. Using the values $\tau_{asy} \sim 0.7$~ps and $\tau_{p}$  $\sim0.3$~ps estimated  from our WAL data (see SI) and reported elsewhere\cite{gorbachev2007weak}, $\lambda_r$ was found to be $85$~meV. As seen from the phase diagram in Fig.~\ref{fig:figure4}(d), this value of $\lambda_r$ is enough to create helical edge-modes in the BLG. 
	
The presence of hBN aligned with the top layer of the BLG (with the consequent moir$\mathrm{\acute{e}}$ superlattice potential) suggests that the observed $R_{NL}$ may also plausibly arise either from Valley Hall (VH) or Quantum Valley Hall (QVH) effect. The first scenario can be ruled out by noting that in VH's case, transport proceeds through the bulk of the material resulting in $R_{NL}$ scaling as $\rho^3$, where $\rho$ is the local 4-probe longitudinal resistivity. Fig.~\ref{fig:figure4}(c) shows that in our device, $R_{NL}$ scales linearly with $\rho$, which is what one expects for charge transport confined to the edges of the device~\cite{PhysRevB.79.035304,shimazaki2015generation}. Note also that the bulk bandgap for BLG at $\left|D\right|=0.2\times{10}^9$~V/m is around 10~meV~\cite{kanayama2015gap}, which ensures that when the Fermi energy is near the center of the bulk bandgap at large $\left|D\right|$, there is no contribution from the bulk of the BLG to electrical transport in the system in the temperature range explored experimentally. In the case of the QVH effect~\cite{li2016gate, Li1149}, resistance quantization has been observed primarily near the CDP, not the PDP. On the contrary, the quantization results presented in this article were all obtained at the PDP. Measurements of $R_{NL}$ in the presence of a perpendicular magnetic field shows that for ${B <B}_{max}$, the non-local signal also appears only near the PDP (see SI).  These, and the fact that the 4-probe conductance quantization observed in our device was $2e^2/h$ (and not $4e^2/h$ expected in the case for QVH for BLG) effectively rule out QVH as the origin of our observations.
	
\section{Conclusions}
	
In summary, the experimental results presented in this article provide an unambiguous demonstration of the TR invariant helical edge-modes in bilayer graphene. We could tune the phase-space with an external displacement field to induce a topological phase transition between a trivial phase and a possible QSHI phase. In the topological phase, the linear conductance was quantized to $2e^2/h$, indicating the presence of a helical edge-mode (a pair of counter-propagating chiral modes). Non-local measurements show that these helical edge-modes are dissipationless and that they equilibrate at the contact probes. Our device gives a  realization of the much sought after moir\'e superlattice with spin-orbit coupling in a mesoscopic system and can form the platform for the realization of the predicted quantum spin-valley hall insulator phase~\cite{PhysRevLett.108.196802}. Additionally, it provides an experimental realization of a non-interacting version of a system which, in the limit of strong interactions, turns into the recently discovered Quantum Spin Liquid~\cite{meng2010quantum}.
	
	\section{Methods}
	\subsection{Experimental methods}The 2-D flakes used in our device structure were obtained by mechanical exfoliation from single-crystals purchased from HQ graphene. The size of the obtained flakes ranged from approximately 100-400~$\mathrm{\mu m^2}$. The exfoliated flakes were stacked on top of each other in the desired sequence using a home build micro-manipulator under an optical microscope and then transferred on to SiO$_2$/Si$^{++}$ wafers. Standard electron beam lithography followed by etching and metal deposition was used to define the electrical contacts. The details of device fabrication are described in SI 1. All electrical measurements were performed in a cryogen-free dilution refrigerator at a base temperature of 20~mK using standard low-frequency lock-in techniques with excitation currents of 1~nA. 	
	
	\subsection{Theoretical calculations}Theory method: The theoretical calculation is done within a material-specific tight-binding model with spin-orbit coupling (SOC) and electric field on the bilayer graphene. The electric field oppositely polarizes the two graphene layers, and that the proximity effects due to the \ch{WSe2} substrate are higher on the bottom layer than in the top layer. To take into account these effects, we consider the onsite potential of the two layers to be different, breaking the sublattice symmetry. The SOC terms include both Ising and Rashba-type terms. The spin-orbit splitting also becomes anisotropic between the conduction and valence bands. Their combined effect is shown to garner an inverted bandgap in the non-trivial topological phase, with a helical edge state as confirmed by edge state calculations in a lattice model with an open boundary condition.

	\begin{acknowledgement}
		The authors acknowledge fruitful discussions with Anindya Das, Atindra Nath Pal, and Hemanta Kumar Kundu. The authors acknowledge device fabrication facilities in NNFC, CeNSE, IISc. S.K.S. acknowledges PMRF, MHRD for financial support. A.B acknowledges funding from SERB (HRR/2015/000017), DST (DST/SJF/PSA-01/2016-17), and IISc. 
		
	\end{acknowledgement}

	\begin{suppinfo}
		
		Supporting information contains (a) details of device fabrication, (b) detailed results of magnetoresistance measurements, (c) details of measurement of quantized conductance in other multi-terminal configurations, and (d) calculations of edge conductance based on Landauer B\"{u}ttiker  formalism.
		
	\end{suppinfo}

\section{Associated Content}

A pre-print version of the article is available at: Priya Tiwari; Saurabh Kumar Srivastav; Sujay Ray; Tanmoy Das and Aveek Bid. Quantum Spin Hall effect in bilayer graphene heterostructures. 2020, arXiv:2003.10292 (cond-mat). ArXiv. https://arxiv.org/abs/2003.10292 (accessed March 23, 2020).
	
	\clearpage
	
	\clearpage

\providecommand{\latin}[1]{#1}
\makeatletter
\providecommand{\doi}
{\begingroup\let\do\@makeother\dospecials
	\catcode`\{=1 \catcode`\}=2 \doi@aux}
\providecommand{\doi@aux}[1]{\endgroup\texttt{#1}}
\makeatother
\providecommand*\mcitethebibliography{\thebibliography}
\csname @ifundefined\endcsname{endmcitethebibliography}
{\let\endmcitethebibliography\endthebibliography}{}

\end{document}